\documentclass[aps,twocolumn,prl,showpacs,preprintnumbers,%
superscriptaddress]{revtex4}
\renewcommand\d{\partial}

\usepackage{amsmath,amssymb,bm,comment}
\usepackage{hyperref}

\def\d{\partial}

\def\and{{\rm and}}

\begin{document}

\preprint{INT-PUB-09-030}
\title{Hydrodynamics with Triangle Anomalies}
\author{Dam T.~Son}
\affiliation{Institute for Nuclear Theory, University
of Washington, Seattle, Washington 98195-1550, USA}
\author{Piotr Sur\'owka}
\affiliation{Department of Physics, University of Washington,
Seattle, Washington 98195-1560, USA}
\affiliation{Institute of Physics, Jagiellonian University, Reymonta
4, 30-059 Krak\'ow, Poland}
\date{June 2009}
\begin{abstract}
We consider the hydrodynamic regime of theories with quantum anomalies
for global currents.  We show that a hitherto discarded term in the
conserve current is not only allowed by symmetries, but is in fact
required by triangle anomalies and the second law of thermodynamics.
This term leads to a number of new effects, one of which is chiral
separation in a rotating fluid at nonzero chemical potential.  The new
kinetic coefficients can be expressed, in a unique fashion, through
the anomalies coefficients and the equation of state.  We briefly
discuss the relevance of this new hydrodynamic term for physical
situations, including heavy ion collisions.
\end{abstract}
\maketitle

\emph{Introduction.}---Relativistic hydrodynamics is important for
many questions in nuclear physics, astrophysics, and cosmology.  For
example, hydrodynamic models are used extensively for describing the
evolution of the fireball created in heavy-ion collisions.  The
relativistic hydrodynamic equations have been proposed many years
ago~\cite{Eckart:1940te,0750627670}; such equations describe the
dynamics of an interacting relativistic theory at large distance and
time scales.  The hydrodynamic variables are the local velocity
$u^\mu(x)$ (satisfying $u^2=-1$), the local temperature $T(x)$ and
chemical potential(s) $\mu^a(x)$, where the index $a$ numerates the
conserved charges.  The hydrodynamic equations govern the time
evolution of these variables; they have the form of the conservation
laws $\d_\mu T^{\mu\nu} = 0$, $\d_\mu j^{a\mu}=0$, supplemented by the
constitutive equations which express $T^{\mu\nu}$ and $j^{a\mu}$ in
terms of $u^\mu$, $T$, and $\mu^a$.  These equations are the
relativistic generalization of the Navier-Stokes equations.

One feature of relativistic quantum field theory that does not have
direct counterpart in nonrelativistic physics is the presence of
triangle anomalies~\cite{Adler:1969gk,Bell:1969ts}.  For currents
associated with global symmetries, the anomalies do not destroy
current conservations, but are reflected in the three-point functions
of the currents.  When the theory put is put in external background
gauge fields coupled to the currents, some of the currents will no
longer be conserved.

In this paper, we show that the presence of quantum triangle anomalies
leads to an important modification of the hydrodynamic equations.  In
other words, in a hot and dense medium quantum anomalies are
expressed macroscopically.  This modification should be important in
many physical situations, including the quark gluon plasma where the
small masses of the $u$ and $d$ quarks can be neglected.

In the simplest case when there is one U(1) current with a U(1)$^3$
anomaly, the constitutive equation for the conserved current $j^\mu$
must contains an additional term proportional to the
vorticity~\footnote{We consider only global currents that are not
coupled to dynamical gauge fields, and assume the associated
symmetries are not spontaneously broken.}
\begin{align}
  j^\mu &= nu^\mu 
  -\sigma T (g^{\mu\nu}+u^\mu u^\nu)\d_\nu\left(\frac\mu T\right)
  + \xi \omega^\mu, \\
  \omega^\mu &=
  \frac12 \epsilon^{\mu\nu\lambda\rho}u_\nu \d_\lambda u_\rho,
\end{align}
where $n$ is the charge density, $\sigma$ is the conductivity, and
$\xi$ is the new kinetic coefficient.

Even in a parity-invariant theory, the vorticity-induced current
$\xi\omega^\mu$ is allowed by symmetries if, e.g., $j^\mu$ is a chiral
current.  This term contains only one spatial derivative, and its
effect is as important as those of viscosity or diffusion.
Nevertheless, this term has been ignored so far.  In fact, if one
follows the standard textbook derivation~\cite{0750627670}, the new
term seems to be disallowed by the existence of an entropy current
with manifestly positive divergence, required by the second law of
thermodynamics.

Recently, however, calculations using the techniques of gauge/gravity
duality~\cite{Maldacena:1997re,Gubser:1998bc,Witten:1998qj} within a
particular model (${\cal N}=4$ super-Yang-Mills plasma with an
R-charge density) gives a nonzero value for
$\xi$~\cite{Erdmenger:2008rm,Banerjee:2008th,Torabian:2009qk}.  
This indicates that
the problem with the entropy current must be circumvented in some way.

In this paper we show that this new term is not only allowed, but is
required by anomalies.  Moreover, the parity-odd kinetic coefficient
$\xi$ is completely determined by the anomaly coefficient $C$, defined
through the divergence of the gauge-invariant current, $\d_\mu j^\mu =
-\tfrac18 C\epsilon^{\mu\nu\alpha\beta}F_{\mu\nu}F_{\alpha\beta}$, and
the equation of state,
\begin{equation}\label{xiC}
  \xi = C \left(\mu^2 - \frac23\frac{\mu^3 n}{\epsilon+P} \right)\!,
\end{equation}
where $\epsilon$ and $P$ are the energy density and pressure.  In the
case of multiple U(1) conserved currents, the formulas are modified
only slightly.  Namely, Eq.~(\ref{xiC}) becomes
\begin{equation}\label{xiC-mult}
%  \xi^a = C^{abc} \left(\mu^b\mu^c 
%    - \frac23\frac{\mu^b\mu^c\mu^d n^d}{\epsilon+P} \right)\!,
  \xi^a = C^{abc} \mu^b\mu^c 
    - \frac23 n^a C^{bcd} \frac{\mu^b\mu^c\mu^d}{\epsilon+P}\,,
\end{equation}
where $a,b,c$ numerate the currents, $C^{abc}$ is symmetric under
permutations of indices and is determined from the anomalies, $\d_\mu
j^{a\mu}=-\tfrac18 C^{abc} \epsilon^{\mu\nu\alpha\beta} F^b_{\mu\nu}
F^c_{\alpha\beta}$.

In the local rest frame of fluid, the new contribution to the current
is ${\bf j} = \tfrac12\xi \bm{\nabla}\times {\bf v}$, which means that
there is a current directed along the vorticity.  For example, in a
volume of rotating quark matter, quarks with opposite helicities will
move to opposite directions, a phenomenon which can be called
hydrodynamic chiral separation.  We note that the effect is present
already at the level of first-order hydrodynamics.  In this respect,
this new term in the current is very unusual: it does not exist in
nonrelativistic solutions of chiral molecules (e.g., sugar) in water,
or in suspensions of chiral objects.  In these cases, the chiral
currents contains terms proportional to the third derivatives of
velocity and terms quadratic in first derivatives of
velocity~\cite{ASS}.  The vorticity-induced current is specific for
relativistic quantum field theory with quantum anomalies.

\emph{Entropy current in hydrodynamics with anomalies}---For
simplicity consider first a relativistic fluid with one conserved
charge, with a U(1)$^3$ anomaly.  To constrain the hydrodynamic
equation, we turn on a slowly varying background gauge field $A_\mu$
coupled to the current $j_\mu$.  We take the strength of $A_\mu$ is of
the same order as the temperature and the chemical potential, so
$A_\mu\sim O(p^0)$ and $F_{\mu\nu}\sim O(p)$.  As in first-order
hydrodynamics, we keep terms of order $O(p)$ in the constitutive
equations for $T^{\mu\nu}$ and $j^\mu$ (or terms of order $O(p^2)$ in
the equations of motion).  Note that $A_\mu$ is not dynamical.

In the presence of an external background field the hydrodynamic
equations obtain the form
\begin{equation}
  \d_\mu T^{\mu\nu}  = F^{\nu\lambda} j_\lambda\,,\quad
  \d_\mu j^\mu = C  E^\mu B_\mu\,.
\end{equation}
where we have defined the electric and magnetic fields in the fluid
rest frame, $E^\mu= F^{\mu\nu} u_\nu$,
$B^\mu=\tfrac12\epsilon^{\mu\nu\alpha\beta}u_\nu F_{\alpha\beta}$.
The right hand sides of these equations take into account the fact
that external field performs work on the system, and the anomaly.
Note that the right hand sides are of order $O(p)$ or $O(p^2)$ in our
power counting, which are within the order hydrodynamic equations.

The stress-energy tensor and the current are
\begin{align}
  T^{\mu\nu} &= (\epsilon+P)u^\mu u^\nu + Pg^{\mu\nu} + \tau^{\mu\nu},\\
  j^\mu &= nu^\mu +\nu^\mu,
\end{align}
where $\tau^{\mu\nu}$ and $\nu^\mu$ are terms of order $O(p)$ which
incorporate, in particular, dissipative effects.  Following Landau and
Lifshitz, we can always require $u_\mu\tau^{\mu\nu}= u_\mu \nu^\mu=0$.
We find $\tau^{\mu\nu}$ and $\nu^\mu$ from the requirement of the
existence of an entropy current $s^\mu$ with non-negative derivative,
$\d_\mu s^\mu\ge0$.  Transforming $u_\nu \d_\mu T^{\mu\nu}+\mu\d_\mu
j^\mu$ using hydrodynamic equations and $\epsilon+P=Ts+\mu n$, we find
\begin{multline}\label{s-old}
  \d_\mu \left(su^\mu-\frac\mu T\nu^\mu\right) =
  -\frac1T \d_\mu u_\nu\tau^{\mu\nu} - \nu^\mu\Bigl(
  \d_\mu\frac\mu T - \frac{E_\mu}T \Bigr)\\
  - C \frac\mu T E\cdot B.
\end{multline}
In the standard treatment when the current is not anomalous, $C=0$,
this equation is interpreted as the the equation of entropy
production.  The first-derivative parts of the energy tensor and the
current have the following form
\begin{align}
  \tau^{\mu\nu} &= -\eta P^{\mu\alpha} P^{\nu\beta}
   (\d_\alpha u_\beta {+}\d_\beta u_\alpha)
   - \bigl(\zeta{-}\tfrac23\eta\bigr) P^{\mu\nu} \d\cdot u,\\
  \nu^\mu &= -\sigma T P^{\mu\nu}\d_\nu \left(\frac\mu T\right) 
            + \sigma E^\mu,
\end{align}
where $P^{\mu\nu}=g^{\mu\nu}+u^\mu u^\nu$, and the entropy production
rate is manifestly positive.  However, in the presence of anomalies
the last term in Eq.~(\ref{s-old}) can have either sign, and can
overwhelm the other terms.  Therefore, the hydrodynamic equations have
to be modified.

The most general modification one can make is to add the following
terms to the U(1) and entropy currents,
\begin{align}
  \nu^\mu &= -\sigma T P^{\mu\nu}\d_\nu \left(\frac\mu T\right) 
            + \sigma E^\mu + \xi \omega^\mu + \xi_B B^\mu,\\
  s^\mu &= su^\mu - \frac\mu T\nu^\mu + D\omega^\mu + D_B B^\mu,
\end{align}
where $\xi$, $\xi_B$, $D$, and $D_B$ are functions of $T$ and $\mu$.
Requiring $\d_\mu s^\mu\ge0$ for an arbitrary initial condition, and
using the following identities which follow from the ideal
hydrodynamic equations,
\begin{align}
  \d_\mu\omega^\mu &= -\frac2{\epsilon+P}\omega^\mu(\d_\mu P- nE_\mu), \\
  \d_\mu B^\mu &= -2\omega\cdot E + \frac1{\epsilon+P}
   (-B\cdot \d P + nE\cdot B),
\end{align}
one finds that the following four equations have to be satisfied
\begin{align}
  & \d_\mu D - 2\frac{\d_\mu P}{\epsilon+P} D - \xi\d_\mu \frac \mu T = 0,
    \label{eq1}\\
  & \d_\mu D_B - \frac{\d_\mu P}{\epsilon+P} D_B - \xi_B\d_\mu 
    \frac \mu T = 0, \label{eq2} \\
  & \frac{2n D}{\epsilon+P}-2D_B + \frac\xi T = 0, \label{eq3} \\
   & \frac{n D_B}{\epsilon+P} + \frac{\xi_B}T - C \frac\mu T =0. \label{eq4}
\end{align}

To proceed further, we change variables from $\mu$, $T$ to a new pair
of variables, $\bar\mu\equiv\mu/T$ and $P$.  From $dP=sdT+nd\mu$, it
is easy to derive
\begin{equation}\label{Maxwell}
  \left(\frac{\d T}{\d P}\right)_{\bar\mu} = \frac T{\epsilon+P}\,, \qquad
   \left(\frac{\d T}{\d \bar\mu}\right)_P = - \frac{nT^2}{\epsilon+P}\,.
\end{equation}
Writing $\d_i D = (\d D/\d P) \d_i P + (\d D/D\bar\mu)\d_i\bar\mu$,
and noting that $\d_i P$ and $\d_i\bar\mu$ can be arbitrary, as they
can be considered as initial condition on a time slice, 
Eq.~(\ref{eq1}) becomes two equations
\begin{equation}\label{eqCD-comp}
   -\xi + \frac{\d D}{\d \bar\mu} = 0, \quad
  \frac{\d D}{\d P}-\frac2{\epsilon+P} D =0.
\end{equation}
Using Eq.~(\ref{Maxwell}), one finds that the most general solution to
Eqs.~(\ref{eqCD-comp}) is
\begin{equation}
  D = T^2d(\bar\mu), 
  \quad \xi = \frac{\d}{\d \bar\mu} \left(T^2 d(\bar\mu)\right)_P ,
%  = T^2 d'(\bar\mu) - \frac{2nT^3}{\epsilon+P} d(\bar\mu)
\end{equation}
where $d(\bar\mu)$ is, for now, an arbitrary function of one variable.
Equation~(\ref{eq2}) yields
\begin{equation}
  D_B = T d_B(\bar\mu),\quad
  \xi_B = \frac{\d}{\d \bar\mu} \left( T d_B(\bar\mu)\right)_P ,
%  = T d_B'(\bar\mu) - \frac{nT^2}{\epsilon+P} d_B(\bar\mu)
\end{equation}
where $d_B(\bar\mu)$ is another function of $\bar\mu$.
From Eqs.~(\ref{eq3}) and (\ref{eq4}) we get
\begin{equation}
  d_B(\bar\mu) = \frac12 d'(\bar\mu), \quad
  d_B'(\bar\mu) - C_{\rm anom} \bar\mu =0,
\end{equation}
which can be integrated.  We find
\begin{equation}
  d_B(\bar\mu) = \frac12 C \bar\mu^2, \quad
  d(\bar\mu) = \frac13 C \bar\mu^3.
\end{equation}
So the new kinetic coefficients are
\begin{equation}\label{xixiB-final}
  \xi = C \left(\mu^2 - \frac23 \frac{n\mu^3}{\epsilon+P}\right) \!,
  \ \
  \xi_B = C \left(\mu-\frac12\frac{n\mu^2}{\epsilon+P}\right)\!.
\end{equation}

\emph{Extension to multiple charges.}---It is easy to extend to the
case of many charges.  Here we consider only the U(1) charges that
commute with each other.  We denote the anomaly coefficients as
$C^{abc}$, which is totally symmetric under permutation of indices and
give the divergence of the gauge-invariant currents,
\begin{equation}
  \d_\mu j^{a\mu} = C^{abc} E^b\cdot B^c.
\end{equation}
The constitutive equations can is now
\begin{equation}
  j^{a\mu} = n^a u^\mu + \cdots + \xi^a\omega^\mu + \xi_B^{ab} B^\mu,
\end{equation}
where $\xi^a$ and $\xi_B^{ab}$ are new transport coefficients.  The
entropy current is now modified to
\begin{equation}
  s^\mu = su^\mu - \frac{\mu^a}T\nu^a + D\omega^\mu + D_B^a B^{a\mu}.
\end{equation}
Repeating the calculations of the previous section, we find that
\begin{equation}
  D = \frac13 C^{abc} T^2 \bar\mu^a\bar\mu^b\bar\mu^c,
  \quad D_B^a = \frac12 C^{abc} T \bar \mu^b\bar\mu^c,
\end{equation}
\begin{equation}
  \xi^a = \frac{\d}{\d\bar\mu^a} D\Big|_P,\qquad
  \xi_B^{ab} = \frac{\d}{\d\bar\mu^a} D^b_B\Big|_P\,,
\end{equation}
and, by using thermodynamic relations, one derives
Eq.~(\ref{xiC-mult}).

\emph{Gravity calculation}---%
The discussion above has been completely independent of details of the
theory.  We would like to check our formulas for the case when the
kinetic coefficients can be calculated explicitly.  In this paper we
use a holographic model as a testing ground for our predictions.
Namely, we look at the theory described by the following 5D action,
\begin{multline}\label{action:eq}
S=\frac{1}{16\pi G_5}\!\int\!d^5x\, \sqrt{-g_5} \Bigl(R + 12 -
F_{AB}F^{AB}\\
+\frac{4\kappa}{3}\epsilon{}^{LABCD}A_L F_{AB} F_{CD}
\Bigr).
\end{multline}
Here Latin indices $A,B$ denote bulk 5D coordinates $r,v,x,y,z$, and
Greek indices $\mu,\nu \in\{v,x,y,z\}$ denote the boundary coordinates
($v$ play the role of time on the boundary).  The above action is a
consistent truncation of type IIB supergravity Lagrangian on
AdS$_5\times$S$^5$ background with a cosmological constant
$\Lambda=-6$ and the Chern-Simons parameter $\kappa=-1/(2\sqrt{3})$
\cite{Cvetic:1999ne,Chamblin:1999tk}.  In this case it describes
${\cal N}=4$ supersymmetric Yang-Mills theory at strong coupling,
where the U(1) charge corresponds to one particular subgroup of SO(6)
internal symmetry.  To keep the discussion general we will keep the
$\kappa$ coefficient unfixed, and treat Eq.~(\ref{action:eq}) as the
definition of our theory.

The field equations corresponding to \eqref{action:eq} are
\begin{align}\label{einmax:eq}
G_{AB}-6 g_{AB} + 2\Bigl(F_{AC}F^C{}_{B}+\tfrac{1}{4} g_{AB} 
F^2\Bigr) &= 0,\\
\nabla_B F^{BA} + \kappa \epsilon^{ABCDE}F_{BC} F_{DE} &=0 \label{einmax:2},
\end{align}
where $g_{AB}$ is the 5D metric, $G_{AB}=R_{AB}-\frac12g_{AB}R$ is the
five dimensional Einstein tensor.  The external gauge field and the
current in the boundary theory are associated with the
asymptotics of the $A_\mu$ near the boundary
\begin{equation}
  A_\mu(r,x) = A_\mu(x) - \frac{2\pi G_5}{r^2} j_\mu(x).
\end{equation}
From Eq.~(\ref{einmax:2})
%the second equation of (\ref{einmax:eq}) 
one derives the relationship
between the anomalies coefficient $C$ and $\kappa$,
\begin{equation}
  C = - \frac2{\pi G_5} \kappa.
\end{equation}

These equations admit an AdS Reissner-Nordstr\"{o}m (RN) black-brane
solution.  In Eddington-Finkelstein coordinates, it is
\begin{align}
  ds^2 &= 2 dv dr - r^2 f(r,m,q) dv^2 + r^2 d{\vec x}^2,\\
  A &= - \frac{\sqrt{3}q}{2r^2} dv,
\end{align}
where
\begin{equation}
  f(r,m,q) = 1 - \frac m{r^4} + \frac{q^2}{r^6}\,.
\end{equation}
The black brane is dual to a fluid at finite temperature $T$ and
chemical potential $\mu$.  The connection between the parameters of
the metric and $T$ and $\mu$ is
\begin{subequations}\label{mqTmu}
\begin{align}
  m &= \frac{\pi^4 T^4}{2^4} (\gamma+1)^3 (3\gamma-1),\ \
  \gamma \equiv \sqrt{1+ \frac{8\mu^2}{3\pi^2 T^2}}\,,\\
  q &= \frac{2\mu}{\sqrt 3} \frac{\pi^2 T^2}4 (\gamma+1)^2.
\end{align}
\end{subequations}

The equation of state of this fluid is
\begin{equation}\label{eos}
  P(T,\mu) = \frac{m(T,\mu)}{16\pi G_5}\, .
\end{equation}

In order to find the hydrodynamic equations, we use the method
developed in Ref.~\cite{Bhattacharyya:2008jc}.  We locally boost the
RN metric and consider the boost velocity $u^\mu$, as well as the mass
and charge of the black hole, as slowly-varying function of the
black-brane coordinates, and also turn on a background gauge field
$A^{\rm bg}_\mu$.  To zeroth order, the background we obtain is
\begin{align}\label{brane:eq}
ds^2&= -2 u_\mu dx^\mu dr   + r^2 (P_{\mu\nu} -f u_\mu u_\nu) dx^\mu dx^\nu, \\
A &= \frac{\sqrt{3} q_0}{2 r^2} u_\mu dx^\mu + A^{\rm bg}_\mu dx^\mu.
\end{align}
By iteration we construct the corrections proportional to first derivatives
\begin{equation}
\begin{split}
g_{AB} &= g^{(0)}_{AB}+ g^{(1)}_{AB}+ \ldots, \\
A_{M} &= A^{(0)}_M+A^{(1)}_M+ \ldots, \\
\end{split}
\end{equation}
requiring the solution to be regular at the horizon.  The solution is
then expanded around the boundary; $\xi$ and $\xi_B$ are read from the
asymptotics of $A_\mu$ near the boundary.  As the result, we
find
%~\cite{Surowka}
\begin{align}
%n\equiv \frac{\sqrt{3} q}{16\pi G_5} \quad;\quad 
\xi &= -\frac{3q^2 \kappa }{2\pi G_5 m} \label{xi-mR}\,, \\
\xi _B & = -\frac{\sqrt{3} \left(3 R^4+m\right) q
  \kappa}{4\pi G_5 m R^2}\,,\label{xiB-mR}
\end{align}
where $R$ is the radius of the horizon, $R=\frac\pi2 T(\gamma{+}1)$.
Equation~(\ref{xi-mR}) is consistent with previous results of
Refs.~\cite{Erdmenger:2008rm,Banerjee:2008th}, while
Eq.~(\ref{xiB-mR}) is a new result. Using the
relationships~(\ref{mqTmu}), one can show that the result coincides
with the previous formulas (\ref{xixiB-final}), computed using the
equation of state~(\ref{eos}).

\emph{Conclusion.}---In this paper we show that the relativistic
hydrodynamic equations have to be modified in order to take into
account effects of anomalies.  At nonzero chemical potentials, we find
a new effect of vorticity-induced current.  Moreover, the kinetic
coefficient characterizing this effect is completely fixed by the
anomalies and the equations of state.

It would be interesting to show that Eqs.~(\ref{xiC}) and
(\ref{xiC-mult}) is valid for all theories with gravity dual.  Perhaps
a proof similar to the one of the constancy of the entropy/viscosity
density ratio~\cite{Buchel:2003tz,Kovtun:2004de} is possible.  It is
also interesting to understand how the new term emerges, for weakly
coupled theories, within the kinetic framework.  Finally, it is
tempting to speculate that Eq.~(\ref{xiC}) can be derived by direct
anomalies matching.  At this moment we do not have a complete
understanding of the microscopic origin of the vorticity-induced
current.

The effect found in this paper is a macroscopic manifestation of one
of the most subtle quantum effects in field theory.  This modification
of the hydrodynamic equation should affect the behavior of a
dense and hot neutrino gas, or of the early Universe with large lepton
chemical potential.  For heavy ion physics, one can draw a parallel
with the ``chiral magnetic effect,'' invoked to explain fluctuations
of charge asymmetry in noncentral
collisions~\cite{Kharzeev:2007jp,Fukushima:2008xe}.

We thank J.~Bhattacharya, M.~Haack, A.~Karch, R.~Loganayagam, and
A.~Yarom for discussions, and H.-U.~Yee for correcting an error
in a previous version of the manuscript. 
DTS is supported, in part, by DOE grant No.\
DE-FG02-00ER41132.  PS is supported, in part, by Polish science grant
NN202 105136 (2009-2011).

\end{document}